# Non-Thermal Radiative Pair Plasmas: Processes and Spectra

Ravi P. Pilla[*] and Jacob Shaham[†]
Department of Physics, Columbia University
New York, N.Y., 10027 U.S.A.

May 16, 1995

## Abstract

We study the emission and absorption spectra due to various photon and pair processes in a non-equilibrium pair plasma containing a significant density of photons. We present here some preliminary results from Monte-Carlo simulations. These investigations are likely to be useful in understanding the radiation and relaxation mechanisms in non-thermal $\gamma$-ray sources in astrophysics.

---

[*]email: ravi@cuphyb.phys.columbia.edu
[†]During the course of this work, Prof. Shaham has tragically passed away. Although he has contributed to the work presented here, he was unable to see the final draft. Therefore, any shortcomings of this manusript should not be attributed to him.

# 1 Introduction

Relativistic plasmas composed of electron-positron pairs, ions, and photons are believed to arise from various astrophysical processes involving neutron stars and black holes. There is some indirect evidence for the existence of such plasmas deduced from observations of active galactic nuclei and various $\gamma$-ray sources. Such plasmas have been studied extensively by many authors in the past. See for example [1 − 3] and references therein. Much of the past research was devoted to the study of equilibrium plasmas with internally generated photons. It will be interesting to study the evolution of a relativistic non-thermal pair plasma with an initial distribution of high-energy $(0.1 − 100$ MeV$)$ photons, whose density is comparable to the pair density, which are not produced by any emission processes within the plasma. This will be of some use in understanding the physics of non-thermal $\gamma$-ray sources in astrophysics. It may also have some relevance to the study of $\gamma$-ray-burst sources in their final stages when they are only moderately optically thick. The present work is an attempt to extend and generalize the existing research on equilibrium pair plasmas mentioned above.

# 2 The Model

We consider a non-equilibrium plasma composed of photons, protons, and relativistic electrons and positrons. The plasma is assumed to be stationary (i.e., confined) and free from any magnetic fields. The number densities of photons, electrons, and positrons are $n_\gamma$, $n_-$, and $n_+$, respectively. They are assumed to be comparable to each other. In such a case many radiation and relaxation mechanisms have to be taken into account to determine the evolution of the plasma. Important photon processes are Compton (and inverse Compton) scattering, double Compton scattering, and pair creation. The particle processes are bremsstrahlung from $e^\pm$-$e^\pm$, $e^\pm$-$e^\mp$, and electron-proton collisions, Møller and Bhabha scattering, and pair annihilation.

## 2.1 The Kinetic Equation

To study the evolution of such a plasma we proceed from the relativistic Boltzmann equation,

$$p^\mu \, \partial_\mu \, f(x,p) = \sum_i [\, \eta_i - \chi_i \, f \,], \qquad (1)$$

where $x^\mu$ and $p^\mu$ are the position and momentum 4-vectors, and $f$ is the distribution function. The sum is over all processes that contribute to emission and absorption, and $\eta_i$ and $\chi_i$ are the corresponding binary coeffiecients. For isotropic emission and absorption, we define the spectral functions for photons and particles,

$$F_\gamma(\varepsilon) = \frac{4\pi\varepsilon^2}{n_\gamma} f_\gamma(\varepsilon) \quad \text{and} \quad F_\pm(\gamma) = \frac{4\pi\gamma\sqrt{\gamma^2-1}}{n_\pm} f_\pm(\gamma), \qquad (2)$$

respectively. Here $\varepsilon$ is the photon energy in units of $mc^2$ and $\gamma$ is the particle Lorentz factor. They are normalized so that

$$\int_0^\infty d\varepsilon\, F_\gamma(\varepsilon) = 1 \quad \text{and} \quad \int_1^\infty d\gamma\, F_\pm(\gamma) = 1. \qquad (3)$$

## 2.2 A Family of Non-Equilibrium Functions

We use the following three-parameter family of non-equilibrium spectral functions:

$$F_\gamma(\varepsilon) = \begin{cases} c_1 \dfrac{\varepsilon^2}{\exp(\varepsilon/\Theta)-1}, & \text{if } 0 < \varepsilon \le \alpha\,\varepsilon_{max}, \\[6pt] c_2\,\varepsilon^{-\delta}, & \text{otherwise.} \end{cases} \qquad (4)$$

Here $\alpha\,\varepsilon_{max}$ represents the spectral break, where $\varepsilon_{max}$ is the energy at which the thermal part has its maximum, and $\Theta = k_B T/m c^2$ is the "temperature" of the low-energy part of the spectrum, where $m$ is the electron rest mass and $k_B$ is the Boltzmann constant. Finally, $\delta$ is the power-law index for the high-energy part. The constants $c_1$ and $c_2$ are determined by the continuity of the function at $\alpha\,\varepsilon_{max}$ and the normalization given by equation (3). A similar family of functions is defined for particles as well.

The motivation for this choice is the following: It has been realized for some time[4] that Møller and Bhabha scattering processes are not efficient enough to maintain particle equilibrium in a relativistic plasma for $\Theta > 3.5$. This result applies to pair-dominated plasmas. However, if we also include energetic photons of comparable density which are non-thermal, then it is likely that a non-thermal tail will develop beyond a few MeV for pairs as well as photons. We parametrize this by a power-law index $\delta$. Besides, similar power-law tails are found in many $\gamma$-ray-burst spectra[5]. The low-energy part is characterized by a fictitious temperature $\Theta$. This is because at low energies many of the relaxation processes are efficient for thermalization. This is only a hypothetical family of functions; we do not imply that they are based on any confirmed observations.

## 2.3 The Absorption and Emission Coefficients

In binary collisions between two species (1 and 2) of particles or photons with spectral functions $F_1$ and $F_2$, the photon emission coefficient is given by,

$$\eta(\varepsilon) = \frac{n_1 n_2\,\varepsilon}{16\pi^2(1+\delta_{12})} \int_R \Big[\prod_{i=1}^2 F_i(\varepsilon_i)d\varepsilon_i\,d\Omega_i\Big] \frac{(p_1\cdot p_2)}{\varepsilon_1\varepsilon_2} \beta_{rel} \frac{d\sigma_{12}}{dP}. \qquad (5)$$

Here $dP = \varepsilon^2 \, d\varepsilon \, d\Omega$ for photons and $\gamma^2 \beta \, d\gamma \, d\Omega$ for particles, $d\sigma_{12}/dP$ is the differential cross section for the process, $\delta_{12} = 1$ for identical particles and is zero otherwise, $c \, \beta_{rel}$ is the relative velocity of the particles, $\varepsilon_i = \varepsilon$ in the case of photons and $\varepsilon_i = \gamma$ in the case of particles, and $p_i$ are the 4-momenta of particles or photons in units of $mc$. The integration is over all initial states, subject to the conservation of 4-momentum so that one of the final states is a photon of energy $\varepsilon$ as required. This usually results in a multi-dimensional integral over a complicated region $R$. The photon absorption coefficient is obtained from a similar formula. In that case the integration is performed over all final states which absorb an initial state of energy $\varepsilon$. Physically, $4\pi \, \varepsilon \eta / mc$ gives the rate at which the photons are emitted per unit volume, per unit time, and per unit energy. An analogous interpretation is given to the photon absorption coefficient and to the particles cases.

## 3   Results and Discussion

So far we have performed Monte-Carlo evaluations of various emissivity integrals for pair annihilation, Compton scattering, and $e^{\pm}$-$e^{\pm}$ and $e^{\pm}$-$e^{\mp}$ bremsstrahlung. To test our calculations we have used relativistic Maxwell-Boltzmann distributions at various temperatures and compared our results on equilibrium plasmas with those of several previous authors[6-9]; our results are found to be in good agreement. Using the same programs, we have calculated the spectra for the family of non-equilibrium functions specified by equation (4). We define a dimensionless emissivity function,

$$\Psi(\varepsilon) = \frac{4\pi\,\varepsilon\,\eta(\varepsilon)}{n_1\,n_2\,r_e^2}, \tag{6}$$

where $r_e$ is the classical electron radius. The function $\Psi$ due to pair annihilation and Compton scattering is shown in Figure 1. We are presently computing the emission and absorption functions for the remaining photon and particle interactions using the same Monte-Carlo scheme. This will enable us to understand various relaxation and radiation mechanisms in such plasmas. We will report these investigations in greater detail elsewhere[10].

## Acknowledgements

This work was supported in part by NASA grant NAG5-2017. RPP thanks M. Kamionkowski for help and many useful comments. This work is contribution number 569 of the Columbia Astrophysics Laboratory.

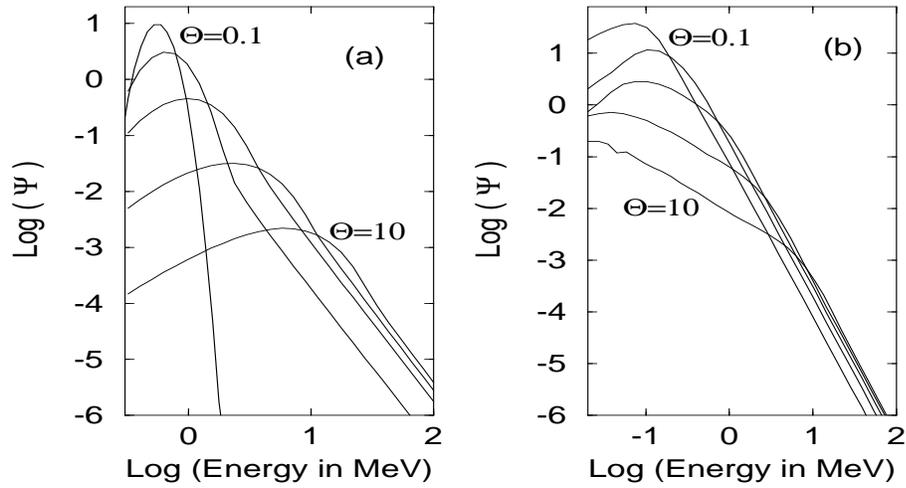

Figure 1: Function $\Psi$ for (a) pair annihilation and (b) Compton scattering. In all cases we use $\alpha = 3$ and $\delta = 2$, $\Theta = 0.1$, 0.33, 1, 3.3 and 10.